\documentclass[11pt]{article}
\usepackage{latexsym} \usepackage{epsf}
\usepackage{a4}
\usepackage{amsfonts}
\usepackage{amsmath}
\usepackage{amsthm}
\usepackage{amssymb}
\usepackage[cp1251]{inputenc}
\usepackage{cite}
\usepackage{verbatim}
\usepackage{units}

\renewcommand{\theequation}{\thesection.\arabic{equation}}
\newcounter{subequation}[equation]
\makeatletter

\expandafter\let\expandafter\reset@font\csname reset@font\endcsname

\def\subeqnarray{\arraycolsep1pt
    \def\@eqnnum\stepcounter##1{\stepcounter{subequation}%
        {\reset@font\rm(\theequation\alph{subequation})}}
\jot5mm     \eqnarray}

\makeatother

\def\be{\begin{equation}}
\def\ee{\end{equation}}
\def\lb{\label}
\def\bea{\begin{eqnarray}}
\def\eea{\end{eqnarray}}

\def\one#1{#1^{\raise5pt\hbox{$\scriptstyle\!\!\!\!1$}}\,{}}
\def\two#1{#1^{\raise5pt\hbox{$\scriptstyle\!\!\!\!2$}}\,{}}

\def\II{\hbox{{1}\kern-.25em\hbox{l}}}
\makeatletter
\def\binrel@#1{\begingroup
  \setboxz@h{\thinmuskip0mu
    \medmuskip\m@ne mu\thickmuskip\@ne mu
    \setbox\tw@\hbox{$#1\m@th$}\kern-\wd\tw@
    ${}#1{}\m@th$}%
  \edef\@tempa{\endgroup\let\noexpand\binrel@@
    \ifdim\wdz@<\z@ \mathbin
    \else\ifdim\wdz@>\z@ \mathrel
    \else \relax\fi\fi}%
  \@tempa
}
\let\binrel@@\relax
\def\overset#1#2{\binrel@{#2}%
  \binrel@@{\mathop{\kern\z@#2}\limits^{#1}}}
\def\underset#1#2{\binrel@{#2}%
  \binrel@@{\mathop{\kern\z@#2}\limits_{#1}}}
\makeatother
\newfont{\bbd}{msbm10 scaled\magstep1}
\def\C{\hbox{\bbd C}}

\begin{document}


\begin{titlepage}

\vspace*{1cm}

\begin{center}
{\LARGE \bf{The $\mathrm{R}$-operator for Modular Double}}

\vspace{1cm}

{\large \sf D. Chicherin$^{ba}$\footnote{\sc e-mail:chicherin@pdmi.ras.ru},
  S. Derkachov$^{a}$\footnote{\sc e-mail:derkach@pdmi.ras.ru}
}

\vspace{0.5cm}

\begin{itemize}
\item[$^a$]
{\it St. Petersburg Department of Steklov Mathematical Institute
of Russian Academy of Sciences,
Fontanka 27, 191023 St. Petersburg, Russia}
\item[$^b$]
{\it Chebyshev Laboratory, St.-Petersburg State University,\\
14th Line, 29b, Saint-Petersburg, 199178 Russia}
\end{itemize}
\end{center}
\vspace{0.5cm}
\begin{abstract}
We construct the R-operator -- solution of the
Yang-Baxter equation acting in the tensor
product $\pi_{s_1}\otimes\pi_{s_2}$ of two infinite-dimensional
representations of Faddeev's modular double.
This R-operator intertwines the product of two L-operators
associated with the modular double and
it is built from three basic operators generating
the permutation group of four parameters $\mathfrak{S}_4$.
\end{abstract}

\vspace{4cm}

\end{titlepage}


{\small \tableofcontents}
\renewcommand{\refname}{References.}
\renewcommand{\thefootnote}{\arabic{footnote}}
\setcounter{footnote}{0} \setcounter{equation}{0}


\section{Introduction}
\setcounter{equation}{0}

We construct the solution of the Yang-Baxter equation
\begin{equation}\label{YB}
\mathbb{R}_{12} (u-v)\,\mathbb{R}_{13}(u)\, \mathbb{R}_{23}(v)
=\mathbb{R}_{23}(v)\,\mathbb{R}_{13}(u)\,\mathbb{R}_{12}(u-v)
\end{equation}
where the operators $\mathbb{R}_{ik}(u)$ are acting in the
tensor product of two infinite-dimensional representations
$\pi_{s_i}\otimes\pi_{s_k}$ of the modular double
of $\mathcal{U}_q(s\ell_2)$.

Firstly, we solve the defining
$\mathrm{RLL}$-relation~\cite{KRS}
\begin{equation}\label{RLL}
\mathbb{R}_{12}(u-v)\,\mathrm{L}_1(u)\,\mathrm{L}_2(v)=
\mathrm{L}_2(v)\,\mathrm{L}_1(u)\,\mathbb{R}_{12}(u-v)\,
\end{equation}
using as the main building blocks two operators.
The first building block is an intertwining operator for the equivalent
representations $\pi_s$ and $\pi_{-s}$, and the second building block is obtained from the intertwining
operator by some duality transformation.

The proof that the obtained R-operator obeys
the general Yang-Baxter equation is based on the Coxeter or
star-triangle relation for two main building blocks.

The operator $\mathbb{R}_{12}(u)$ was
constructed in the work of A.Bytsko and J.Teschner~\cite{BT1}
as a function of the operator argument.
We hope that our construction is simpler and
allows generalization to a higher rank~\cite{FIp,Ip}.
It shows that R-operator can be represented
in two forms: as a product of four simple and explicit
operators and as an integral operator.

In the last Section we consider some reductions of
the obtained operator $\mathbb{R}_{12}(u)$ and
reproduce the universal R-operator for the modular double from~\cite{F,BT}.

\section{Modular double and intertwining operator}
\setcounter{equation}{0}

\subsection{Modular double}

We consider the modular double of $\mathcal{U}_q(s\ell_2)$ introduced by L.D.Faddeev in~\cite{F}.
This algebra is formed by two sets of generators
$\mathbf{E}\,,\mathbf{K}\,,\mathbf{F}$ and
$\widetilde{\mathbf{E}}\,,\widetilde{\mathbf{F}}\,,\widetilde{\mathbf{K}}$.
The usual relations for $\mathbf{E}\,,\mathbf{K}\,,\mathbf{F}$
\begin{equation} \label{qsl2}
\begin{array}{c}
[\mathbf{E},\mathbf{F}] = \frac{\mathbf{K}^2 - \mathbf{K}^{-2}}{q-q^{-1}} \;,\;\;\;
\mathbf{K} \mathbf{E} = q \mathbf{E} \mathbf{K} \;,\;\;\;
\mathbf{K} \mathbf{F} = q^{-1} \mathbf{F} \mathbf{K}\,,
\end{array}
\end{equation}
where $q = e^{i \pi \tau}$ ($\tau \in \C$ and it is not a rational number),
are supplemented by similar relations for $\widetilde{\mathbf{E}},\,\widetilde{\mathbf{F}},\,\widetilde{\mathbf{K}}$
with parameter $\widetilde{q} = e^{i \pi / \tau}$. The generators $\mathbf{E}$,$\mathbf{F}$ commute with $\widetilde{\mathbf{E}}$,$\widetilde{\mathbf{F}}$. The generator
$\mathbf{K}$ anti-commutes with $\widetilde{\mathbf{E}}$,$\widetilde{\mathbf{F}}$ and $\widetilde{\mathbf{K}}$ anti-commutes with $\mathbf{E}$,$\mathbf{F}$.
This algebra possesses two central elements -- Casimir operators. One of them has the form
\begin{equation} \label{Casimir}
\mathbf{C}= \left(q-q^{-1}\right)^2\,\mathbf{F} \mathbf{E} - q \mathbf{K}^2 - q^{-1} \mathbf{K}^{-2} + 2\,,
\end{equation}
and the second is constructed out of $\widetilde{\mathbf{E}},\widetilde{\mathbf{F}},\widetilde{\mathbf{K}}$
and $\widetilde{q}$ and its explicit expression is similar to (\ref{Casimir}).

We shall use the parametrization $\tau = \frac{\omega'}{\omega}$ where
the complex numbers $\omega$ and $\omega'$ with positive imaginary parts are restricted by relation $\omega \omega' = -\frac{1}{4}$. Then
$$
q = \exp\left(i \pi \omega' / \omega \right) \;,\;\;\;
\widetilde{q} = \exp\left(i \pi \omega / \omega' \right)\,,
$$
and the change $q \rightleftarrows \widetilde{q}$ is equivalent to $\omega \rightleftarrows \omega'$.

The representations of the modular double, introduced in~\cite{F},
were investigated in~\cite{PT,KLS,BT}.
In the following we deal with representation $\pi_s$ of the modular double generators
by finite-difference operators
$\mathbf{K}_s = \pi_s(\mathbf{K})\,,
\mathbf{E}_s = \pi_s(\mathbf{E})\,,
\mathbf{F}_s = \pi_s(\mathbf{F})$
in the space of entire functions rapidly decaying at infinity along contours parallel to the real line.
It is parameterized by one parameter $s$ which we refer to as a {\it spin},
and generators are given by the explicit formulae~\cite{BT,BT1}
\begin{equation} \label{Gs}
\mathbf{K}_s = e^{-\frac{i \pi}{2\omega} p} \;\;\;,\;\;\;\;
\begin{array}{l}
(q-q^{-1})\mathbf{E}_s \equiv \mathbf{e}_s =
e^{\frac{i \pi x}{\omega}} \left[
e^{-\frac{i \pi}{2 \omega}\left(p -s - \omega''\right)} -
e^{\frac{i \pi}{2 \omega}\left(p -s - \omega''\right)}
\right] \;,\\[0.3 cm]
(q-q^{-1})\mathbf{F}_s \equiv \mathbf{f}_s =
e^{-\frac{i \pi x}{\omega}} \left[
e^{\frac{i \pi}{2 \omega}\left(p + s + \omega''\right)} -
e^{-\frac{i \pi}{2 \omega}\left(p + s + \omega''\right)}
\right]\,, 
\end{array}
\end{equation}
where $p$ denotes momentum operator in coordinate
representation $p = \frac{1}{2 \pi i}\, \partial_{x}$.
The formulae for generators
$\widetilde{\mathbf{K}}_s\,,
\widetilde{\mathbf{E}}_s\,,
\widetilde{\mathbf{F}}_s$ are
obtained by a simple interchange $\omega \rightleftarrows \omega'$.
In the representation $\pi_s$ Casimir operators (\ref{Casimir}) take the form
\be \lb{Casimirs}
\mathbf{C}_s= 4 \cos^2\left( \frac{\pi s}{2 \omega} \right)\;,\;\;\;
\widetilde{\mathbf{C}}_s= 4 \cos^2\left( \frac{\pi s}{2 \omega'} \right)\,.
\ee
The formula (\ref{Casimirs}) implies that
representations $\pi_s$ and $\pi_{-s}$ are equivalent.
In order to show this explicitly, we construct the corresponding intertwining operator $\mathrm{W}$ as a solution of the defining equations
\be \label{intw1}
\mathrm{W}\,\mathbf{K}_s = \mathbf{K}_{-s}\,\mathrm{W}\ \,,\
\mathrm{W}\,\mathbf{E}_s = \mathbf{E}_{-s}\,\mathrm{W}\ \,,\
\mathrm{W}\,\mathbf{F}_s = \mathbf{F}_{-s}\,\mathrm{W}
\ee
and similar equations with generators
$\widetilde{\mathbf{K}}_s\,,\widetilde{\mathbf{E}}_s\,,
\widetilde{\mathbf{F}}_s$.
The explicit construction of intertwining operator
is given in~\cite{PT} and we rederive it here for convenience.
Intertwining operator $\mathrm{W}$ will be a basic building
block in construction of the general $\mathrm{R}$-operator which solves the Yang-Baxter relation (\ref{YB}).

The defining system (\ref{intw1}) is equivalent to a set of functional relations which fix the intertwining operator $\mathrm{\mathrm{W}}$ unambiguously.
The relations
$\mathrm{W}\,\mathbf{K}_s = \mathbf{K}_{-s}\,\mathrm{W}$,
$\mathrm{W}\,\widetilde{\mathbf{K}}_s = \widetilde{\mathbf{K}}_{-s} \, \mathrm{W}$ imply that
the intertwining operator is a function of momentum operator, $\mathrm{W} = \mathrm{W}(p)$.
Then relations
$\mathrm{W}\, \mathbf{E}_s = \mathbf{E}_{-s} \,\mathrm{W}$,
$\mathrm{W}\, \mathbf{F}_{s} = \mathbf{F}_{-s} \,\mathrm{W}$ and their dual lead
to the finite-difference equations
\begin{equation} \label{FunEq'}
\frac{\mathrm{W}(p-\omega')}{\mathrm{W}(p + \omega')} =
\frac{\cos \frac{\pi}{2 \omega} ( p+s)}{\cos \frac{\pi}{2 \omega} ( p-s)}\ \ \,,\ \ \frac{\mathrm{W}(p-\omega)}{\mathrm{W}(p + \omega)} =
\frac{\cos \frac{\pi}{2 \omega'} ( p+s)}{\cos \frac{\pi}{2 \omega'} ( p-s)}\,.
\end{equation}
The solution of these finite-difference equations is given in terms of
some special function.

\subsection{Special functions}

We shall use two basic special functions. The first one is non-compact quantum dilogarithm
which has the following integral representation
\be \lb{gamma}
\gamma(z) = \exp\left(-\frac{1}{4}\int\limits^{+\infty}_{-\infty}
\frac{\mathrm{d}\,t}{t}\,\frac{e^{i t z}}{\sin(\omega t)\sin(\omega^{\prime} t)}\right)\,,
\ee
where the contour goes above the singularity at $t = 0$.
This function is closely related to double sine function of Barnes~\cite{B}, and in the context of quantum integrable systems
L.D.Faddeev pointed out in~\cite{F1} its remarkable properties.
The key formulae for $\gamma(z)$ are given in~\cite{K,FKV,V}.

We shall use notations
\be \lb{nota}
\omega'' = \omega + \omega'
\;,\;\;\; \beta = \frac{\pi}{12}\left( \frac{\omega}{\omega'} +
\frac{\omega'}{\omega} \right)\,.
\ee
The function $\gamma(z)$ respects a pair of finite-difference equations
$$
\frac{\gamma(z+\omega^{\prime})}
{\gamma(z-\omega^{\prime})} = 1+e^{-\frac{i\pi}{\omega}\,z} \ ;\
\frac{\gamma(z+\omega)}{\gamma(z-\omega)} = 1+e^{-\frac{i\pi}{\omega^{\prime}}\,z}\,
$$
and reflection relation
\be \lb{refl}
\gamma(z)\,\gamma(-z) = e^{i\beta}\,e^{i\pi z^2}\,.
\ee
The second function is
\be \lb{D}
D_{a}(z) = e^{-2 \pi i a z} \frac{\gamma(z+a)}{\gamma(z-a)}\,.
\ee
In fact it coincides with the Faddeev-Volkov's $\mathrm{R}$-matrix~\cite{VF,BMS},
and it is extensively used in the paper~\cite{BT1}.
The function $D_{a}(z)$ is even and it obeys a simple reflection relation
\be \lb{Dev}
D_a(z) = D_a(-z) \;\;;\;\; D_a(z) D_{-a}(z) = 1
\ee
and a pair of finite-difference equations
\be \lb{FunEq}
\frac{D_a(z-\omega')}{D_a(z + \omega')} =
\frac{\cos \frac{\pi}{2 \omega} ( z-a)}{\cos \frac{\pi}{2 \omega} ( z+a)} \;\; ; \;\;
\frac{D_a(z-\omega)}{D_a(z + \omega)} =
\frac{\cos \frac{\pi}{2 \omega'} ( z-a)}{\cos \frac{\pi}{2 \omega'} ( z+a)}\,.
\ee
Note that the functions $\gamma(z)$ and $D_{a}(z)$ are invariant at $\omega \rightleftarrows \omega'$.

Now we return to the equations (\ref{FunEq'}) for the intertwining operator.
Comparing (\ref{FunEq'}) with (\ref{FunEq}) one concludes that
\be \lb{Wdiff}
\mathrm{W}(p) = D_{-s}(p)\,.
\ee
Normalization constant in (\ref{Wdiff}) has been chosen such that
$\mathrm{W}^{-1}(p) = D_{s}(p)$ due to (\ref{Dev}).
Using formula for the Fourier transformation~\cite{BT1,K,FKV,V}
\be \lb{FourierD}
A(a) \int\limits^{+\infty}_{-\infty} \mathrm{d}t \; e^{2 \pi i t z } D_{a}(t) =
D_{-\omega''-a}(z)\ \ ;\ \
A(a+\omega'') \equiv \gamma(\omega''-2a)\,e^{-\frac{i \pi}{2}(2a-\omega'')^2-\frac{i \pi}{2}\beta}\,,
\ee
it is possible to represent $\mathrm{W}$ as an integral operator. Indeed, due to (\ref{Wdiff}) and (\ref{FourierD}) one has
\begin{equation} \label{Wint}
\mathrm{W}\,\Phi(x) =
A(s) \int\limits^{+\infty}_{-\infty}\mathrm{d} x' \; D_{s-\omega''}(-x') \; e^{2 \pi i x' p} \; \Phi(x) =
A(s) \int\limits^{+\infty}_{-\infty}\mathrm{d} x' \; D_{s-\omega''}(x-x') \; \Phi(x')\,.
\end{equation}

\section{$\mathrm{L}$-operator and its factorization}

The $\mathrm{L}$-operator is constructed out of
generators of the modular double taken in the representation $\pi_s$ (\ref{Gs})
and has the form~\cite{BT1}
\begin{equation} \label{LBT07}
\mathrm{L}(u) = \left(
\begin{array}{cc}
e^{\frac{i \pi}{\omega} u} \mathbf{K}_s - e^{-\frac{i \pi}{\omega} u} \mathbf{K}^{-1}_s \;\; & \;\; \mathbf{f}_s\\[0.3 cm]
\mathbf{e}_s \;\; & \;\; e^{\frac{i \pi}{\omega} u} \mathbf{K}^{-1}_s - e^{-\frac{i \pi}{\omega} u} \mathbf{K}_s
\end{array} \right)\,.
\end{equation}
The $\mathrm{L}$-operator respects standard intertwining relation
with $4 \times 4$ trigonometric $\mathrm{R}$-matrix which is equivalent to the set of commutation relations~(\ref{qsl2}).
The second $\mathrm{L}$-operator is obtained from
$\mathrm{L}(u)$ by the interchange
$\omega \rightleftarrows \omega'$:
$\widetilde{\mathrm{L}} (u) =
\left. \mathrm{L} (u) \right|_{\omega \rightleftarrows \omega'}$
In the following we indicate formulae only for the $\mathrm{L}$-operator (\ref{LBT07}), and all relations for the $\widetilde{\mathrm{L}}$-operator have the same form with $\omega \rightleftarrows \omega'$.

The $\mathrm{L}$-operator (\ref{LBT07})
can be represented in the factorized form
$$
\mathrm{L}(u_1,u_2)
= \begin{pmatrix}
U_2 & - U_2^{-1} \\
- U_2^{-1}\,e^{\frac{i\pi}{\omega}x} & U_2\,e^{\frac{i\pi}{\omega}x}
\end{pmatrix}
\begin{pmatrix}
e^{-\frac{i\pi }{2\omega}(p-\omega'')} & 0 \\
0 & e^{\frac{i\pi }{2\omega}(p-\omega'')}
\end{pmatrix}
\begin{pmatrix}
-U_1 & U_1^{-1}\,e^{-\frac{i\pi}{\omega}x} \\
- U_1^{-1} & U_1\,e^{-\frac{i\pi}{\omega}x}
\end{pmatrix}
$$
$$
U_1 = e^{\frac{i\pi}{2 \omega}u_1}\ \,,\ U_2 = e^{\frac{i\pi}{2 \omega}u_2}
$$
where we introduced parameters $u_1$ and $u_2$ instead of
$u$ and $s$
\be \lb{lambdaparam}
u_1 = u + \frac{s}{2} + \frac{\omega}{2} - \frac{\omega'}{2} \;;\;\;\;
u_2 = u - \frac{s}{2} + \frac{\omega}{2} - \frac{\omega'}{2}\,.
\ee
Note that in the notation $\mathrm{L}(u)$ we omit for simplicity
the dependence on spin parameter $s$. Below we shall use
notation $\mathrm{L}(u_1,u_2)$ to show all parameters explicitly and the shorthand notations for the building blocks in factorized representation
\be \lb{LBT07Fact}
\mathrm{L}(u_1,u_2) =
M_{u_2} (x) \, H(p) \, N_{u_1}(x)
\ee
where $H(p) = diag\left(e^{-\frac{i\pi }{2\omega}(p-\omega'')}\,,
e^{\frac{i\pi }{2\omega}(p-\omega'')}\right)$ and
$$
M_u(x) = \begin{pmatrix}
U & - U^{-1} \\
- U^{-1}\,e^{\frac{i\pi}{\omega}x} & U\,e^{\frac{i\pi}{\omega}x}
\end{pmatrix}\ \,,\ \
N_u(x) = \left(U^{-2}-U^2\right) M^{-1}_u(x) = \begin{pmatrix}
-U & U^{-1}\,e^{-\frac{i\pi}{\omega}x} \\
- U^{-1} & U\,e^{-\frac{i\pi}{\omega}x}
\end{pmatrix}\,.
$$

\subsection{$\mathrm{L}^{\pm}$-operators}

Taking one of the parameters to infinity we obtain reduced $\mathrm{L}$-operators
$$
\begin{array}{c}
e^{-\frac{i\pi}{2 \omega} u_2}\, \mathrm{L}(u_1,u_2) \to \mathrm{L}^{+}(u_1) \equiv
\left(
\begin{array}{cc}
1 & 0 \\
0 & e^{\frac{i \pi}{\omega} x}
\end{array}
\right) H(p) \, N_{u_1}(x) \;\;\;\; \text{at} \;\; u_2 \to +\infty\,,
\\ [0.2 cm]
e^{-\frac{i\pi}{2 \omega} u_1}\, \mathrm{L}(u_1,u_2) \to \mathrm{L}^{-}(u_2) \equiv
M_{u_2}(x) \, H(p)
\left(
\begin{array}{cc}
-1 & 0 \\
0 & e^{-\frac{i \pi}{\omega} x}
\end{array}
\right) \;\;\;\; \text{at} \;\; u_1 \to +\infty
\end{array}
$$
or explicitly
$$
\mathrm{L}^{+}(u) =
\begin{pmatrix}
e^{-\frac{i\pi }{2\omega}(p-\omega'')} & 0 \\
0 & e^{\frac{i \pi}{\omega} x}\,e^{\frac{i\pi }{2\omega}(p-\omega'')}
\end{pmatrix}
\begin{pmatrix}
-U & U^{-1}\,e^{-\frac{i\pi}{\omega}x} \\
- U^{-1} & U\,e^{-\frac{i\pi}{\omega}x}
\end{pmatrix}\,,
$$
$$
\mathrm{L}^{-}(u) =
\begin{pmatrix}
U & - U^{-1} \\
- U^{-1}\,e^{\frac{i\pi}{\omega}x} & U\,e^{\frac{i\pi}{\omega}x}
\end{pmatrix}
\begin{pmatrix}
-e^{-\frac{i\pi }{2\omega}(p-\omega'')} & 0 \\
0 & e^{\frac{i\pi }{2\omega}(p-\omega'')}\,e^{-\frac{i \pi}{\omega} x}
\end{pmatrix}\,.
$$
Initial operator $\mathrm{L}(u)$ and operators
$\mathrm{L}^{\pm}(u)$ respect the same intertwining relations with trigonometric $\mathrm{R}$-matrix.
Using formulae
$$
e^{-i \pi p^2} e^{\pm\frac{i \pi}{\omega}x} e^{i \pi p^2} =
e^{\pm\frac{i \pi}{\omega}x} e^{\frac{i \pi}{\omega}(\mp p+\omega')}\,,
$$
one can check that $\mathrm{L}^{-}(u)$ and $\mathrm{L}^{+}(u)$ are
unitarily equivalent
\be \lb{L+toL-}
e^{-i \pi p^2}\,\mathrm{L}^{+}(u)\,e^{i \pi p^2} = \mathrm{L}^{-}(u)\,.
\ee
It is possible to reconstruct the initial $\mathrm{L}$-operator in terms of  $\mathrm{L}^{\pm}$-operators
\be \lb{fL-L+}
-e^{\frac{i \pi}{2 \omega}\omega''} \, \mathrm{L}_1(u,v) =
e^{-i \pi (p_2 - x_{12})^2} \, \mathrm{L}^{-}_1(v)\,
\mathrm{L}^{+}_2(u) \,e^{i \pi (p_2 - x_{12})^2}\,\left(
\begin{array}{cc}
e^{\frac{i \pi}{2 \omega} p_2} & 0 \\
0 & e^{-\frac{i \pi}{2 \omega} p_2}
\end{array}
\right)\,,
\ee
\be \lb{fL+L-}
-e^{\frac{i \pi}{2 \omega}\omega''}
\mathrm{L}_2(u,v) = \left(
\begin{array}{cc}
e^{\frac{i \pi}{2 \omega} p_1} & 0 \\
0 & e^{-\frac{i \pi}{2 \omega} p_1}
\end{array}
\right)\,e^{i \pi (p_1 - x_{12})^2} \, \mathrm{L}^{-}_1(v)\,
\mathrm{L}^{+}_2(u) \,e^{-i \pi (p_1 - x_{12})^2}\,,
\ee
where the operators $x,p$ entering
$\mathrm{L}_k\,,\mathrm{L}^{\pm}_k$ are replaced by $x_k\,,p_k$ and $x_{12} = x_1-x_2$.
The proof is straightforward and it is reduced to a simple direct check using (\ref{LBT07Fact}).
This variant of factorization for the Lax-operator is an analog of the factorization obtained
in the context of the chiral Potts model~\cite{BS,KMN,BKMS,DJMM,T}, and it expresses the equivalence of the
corresponding representations of the $\mathrm{RLL}$-algebra~\cite{T1}.

Due to relation~(\ref{L+toL-}) it is possible to use
only one reduced operator as a building block.
This leads to yet another variant of
factorization~\cite{VF1}.
We rewrite (\ref{fL-L+}) taking into account (\ref{L+toL-})
$$
-e^{-\frac{i \pi}{2 \omega}\omega''} \,
\hat{\mathrm{L}}_1(u,v) =
\mathrm{L}^{+}_1(v)\,
\mathrm{L}^{+}_2(u)
\left(
\begin{array}{cc}
e^{\frac{i \pi}{2 \omega} (p_1+x_{12})} & 0 \\
0 & e^{-\frac{i \pi}{2 \omega} (p_1+x_{12})}
\end{array}
\right)
$$
where $\hat{\mathrm{L}}_1(u,v)$ is obtained from $\mathrm{L}_1(u,v)$
by canonical transformation
$$
\begin{array}{lll}
x_1 & \to & X_1 \equiv p_1+x_1 \\ [0.2 cm]
p_1 & \to & P_1  \equiv -x_1+p_2+x_2\,.
\end{array}
$$
Since $p_1+x_{12}$ commutes with $X_1$ and $P_1$ we can impose the constraint $p_1+x_{12} = 0$  that reproduces the
result of~\cite{VF1}.

Let us note that the reduced operators $\mathrm{L}^{\pm}$ are rather well known~\cite{G,V0,BS}.
In order to render them into the form used in the literature,
$$
\mathrm{L}^{+}(u) = -e^{\frac{i\pi}{2\omega}(u+\omega'')}
\begin{pmatrix}
\mathbf{\hat{u}} & -e^{\lambda}\,\mathbf{\hat{v}}^{-1}\\
e^{\lambda}\,\mathbf{\hat{v}} & \mathbf{\hat{u}}^{-1}
\end{pmatrix}\;,\;\;\; \lambda \equiv -\frac{i \pi}{2 \omega} (2u+\omega'')\,,
$$
we need a Weyl pair $\mathbf{\hat{u}},\mathbf{\hat{v}}$,
$$
\mathbf{\hat{u}} = e^{-\frac{i \pi}{2\omega}p}\;,\;\;\;
\mathbf{\hat{v}} = e^{\frac{i \pi}{\omega}x} e^{\frac{i \pi}{2\omega}(p-\omega'')}\;,\;\;\;
\mathbf{\hat{u}}\,\mathbf{\hat{v}} = q\,\mathbf{\hat{v}}\,\mathbf{\hat{u}}\,.
$$


\section{Basic intertwining relations}

Note that interchange $u_1 \rightleftarrows u_2$ is equivalent to the
substitution $s \to -s$. $\mathrm{L}$-operators (\ref{LBT07}) are
linear in generators $\mathbf{E}_s\,,\mathbf{K}_s\,,\mathbf{F}_s$ and
$\widetilde{\mathbf{E}}_s\,,\widetilde{\mathbf{F}}_s\,,\widetilde{\mathbf{K}}_s$
so that the set of intertwining relations (\ref{intw1}) is equivalent to a single matrix relation
\be \lb{WL2}
D_{u_2-u_1}(p)\,\mathrm{L}(u_1,u_2) =
\mathrm{L}(u_2,u_1)\,D_{u_2-u_1}(p)\,
\ee
or, more explicitly, taking into account factorization relation (\ref{LBT07Fact})
\be \lb{WLfact}
D_{u_2-u_1}(p)\, M_{u_2}(x) \, H(p) \, N_{u_1}(x) =
M_{u_1}(x) \, H(p) \, N_{u_2}(x) \,D_{u_2-u_1}(p)\,.
\ee
In the following we indicate formulae only for the $\mathrm{L}$-operator (\ref{LBT07}) and all relations for the $\widetilde{\mathrm{L}}$-operator have the same form with $\omega \rightleftarrows \omega'$.

\subsection{Duality transformation}

To proceed further let us note that
transformation $p \to -x$ and $x \to p$ preserves
canonical commutation relation. Consequently, applying this
transformation to generators in the representation $\pi_s$ (\ref{Gs}) one finds that they remain to fulfil
commutation relations of the modular double (\ref{qsl2}). We refer to corresponding
$\mathrm{L}$-operator as {\it dual}.
The intertwining relation (\ref{WLfact})
subjected to this duality transformation takes the form
\be \lb{WLdual}
D_{u_2-u_1}(x)\cdot M_{u_2}(p)\,H(-x)\,N_{u_1}(p) =
M_{u_1}(p)\,H(-x)\,N_{u_2}(p) \cdot D_{u_2-u_1}(x)
\ee
where we used that $D_{u_2-u_1}(-x)=D_{u_2-u_1}(x)$ (see (\ref{Dev})).
This observation allows us to prove the second intertwining relation
\be \lb{SLL}
D_{u_1-v_2}(x_{12})\,
\mathrm{L}_1(u_1,u_2)\,
\mathrm{L}_2(v_1,v_2) =
\mathrm{L}_1(v_2,u_2)\, \mathrm{L}_2(v_1,u_1)
\,D_{u_1-v_2}(x_{12})\,
\ee
which will be used in the next Section to construct the general $\mathrm{R}$-operator.

\vspace{1 em}

Indeed, similarly to (\ref{WLdual}) the transformation
\be \lb{dual}
p \to x_{21}\equiv x_2-x_1 \ \,,\ \ x \to p_1
\ee
of the relation (\ref{WLfact}) leads to intertwining relation
\be \lb{WLdual2}
D_{u_1-v_2}(x_{12})\, M_{u_1}(p_1)\,H(x_{21})\,N_{v_2}(p_1) =
M_{v_2}(p_1)\,H(x_{21})\,N_{u_1}(p_1) \,D_{u_1-v_2}(x_{12})\,.
\ee
Further we note that dual $\mathrm{L}$-operator in the previous formula can be factorized as follows
$$
M_{u_1}(p_1)\,H(x_{21})\,N_{v_2}(p_1) =
-i e^{\frac{-i \pi }{2 \omega} x_{21}}
H(p_1)\,N_{u_1}(x_1)\,M_{v_2}(x_2)\,H^{-1}(p_1)
\begin{pmatrix}
- e^{\frac{i \pi \omega'}{2 \omega}} & 0 \\
0 & e^{-\frac{i \pi \omega'}{2 \omega}}
\end{pmatrix}\,.
$$
Substituting the latter formula in (\ref{WLdual2}) and
taking into account that $D_{u_1-v_2}(x_{12})$ commutes with
$p_1+p_2$ so that
$$
H(p_2)\,D_{u_1-v_2}(x_{12})\,H^{-1}(p_2) = H(p_1)\,D_{u_1-v_2}(x_{12})\,H^{-1}(p_1)
$$
one obtains
$$
D_{u_1-v_2}(x_{12})\,H(p_1)\, N_{u_1}(x_1)\,M_{v_2}(x_2)\,H(p_2) =
H(p_1)\,N_{v_2}(x_1)\,M_{u_1}(x_2)\,H(p_2)\,D_{u_1-v_2}(x_{12})\,.
$$
This relation is equivalent to (\ref{SLL}) in view of factorization relation (\ref{LBT07Fact}).

\subsection{Basic intertwining relations for $\mathrm{L}^{\pm}$-operators}

We take into account local factorization (\ref{fL-L+}) and rewrite
intertwining relation (\ref{WL2}) using formula $e^{i \pi (p_2 - x_{12})^2} p_1 e^{-i \pi (p_2 - x_{12})^2} = p_1 + p_2 - x_{12}$ in the following form
\be \lb{intwL-L+}
D_{v-u}(p_1+p_2-x_{12})\,\mathrm{L}_1^{-}(v)\,\mathrm{L}_2^{+}(u) =
\mathrm{L}_1^{-}(u)\,\mathrm{L}_2^{+}(v)\,D_{v-u}(p_1+p_2-x_{12})\,.
\ee
Relation between two reduced $\mathrm{L}$-operators (\ref{L+toL-}) allows to rewrite it in another form
\be \lb{intwL+L-}
D_{v-u}(x_{12})\,\mathrm{L}_1^{+}(v)\,\mathrm{L}_2^{-}(u) =
\mathrm{L}_1^{+}(u)\,\mathrm{L}_2^{-}(v)\,D_{v-u}(x_{12})\,,
\ee
and also
$$
\begin{array}{c}
D_{v-u}(p_2-x_{12})\,\mathrm{L}_1^{+}(v)\,\mathrm{L}_2^{+}(u) =
\mathrm{L}_1^{+}(u)\,\mathrm{L}_2^{+}(v)\,D_{v-u}(p_2-x_{12})\,,
\\[0.2 cm]
D_{v-u}(p_1-x_{12})\,\mathrm{L}_1^{-}(v)\,\mathrm{L}_2^{-}(u) =
\mathrm{L}_1^{-}(u)\,\mathrm{L}_2^{-}(v)\,D_{v-u}(p_1-x_{12})\,.
\end{array}
$$

\section{The $\mathrm{R}$-operator}

Now we proceed to construct the general $\mathrm{R}$-operator acting in the tensor product of two representations $\pi_{s_1}\otimes \pi_{s_2}$ (\ref{Gs}).
To do it we solve the $\mathrm{R}\mathrm{LL}$-relation~(\ref{RLL}).
It is convenient to extract from the R-matrix the permutation operator
$\mathbb{R}_{12}(u) = \mathbb{P}_{12}\,\mathrm{R}_{12}(u)$, where
the permutation operator interchanges arguments,
$\mathbb{P}_{12}\,\Phi(z_1,z_2)=\Phi(z_2,z_1)$.
Then we obtain the $\mathrm{R}\mathrm{LL}$-relation with $\mathrm{L}$-operator (\ref{LBT07})
\begin{equation} \label{RLL1}
\mathrm{R}(\mathbf{u}) \,\mathrm{L}_1(u_1,u_2)\,
\mathrm{L}_2(v_1,v_2) =
\mathrm{L}_1(v_1,v_2)\, \mathrm{L}_2(u_1,u_2)\, \mathrm{R}(\mathbf{u})
\end{equation}
and the analogous one for the second $\widetilde{\mathrm{L}}$-operator
\begin{equation} \label{RLL2}
\mathrm{R}(\mathbf{u})\, \widetilde{\mathrm{L}}_1(u_1,u_2)\,
\widetilde{\mathrm{L}}_2(v_1,v_2) =
\widetilde{\mathrm{L}}_1(v_1,v_2) \,\widetilde{\mathrm{L}}_2(u_1,u_2)\,
\mathrm{R}(\mathbf{u})\,.
\end{equation}
Here subscripts $1,2$ in $\mathrm{L}$-operators denote distinct quantum spaces in $\pi_{s_1}\otimes \pi_{s_2}$ where corresponding operators act
nontrivially. The spin parameters $s_1$, $s_2$ and spectral parameters $u$, $v$ are
related with four parameters appearing in the $\mathrm{R}\mathrm{LL}$-relation according to (\ref{lambdaparam}).
We  combine these parameters in one set in the following order
$\mathbf{u}\equiv(u_2,u_1,v_2,v_1)$ and use notation $\mathrm{R}(\mathbf{u})$ to show explicitly the dependence of the $\mathrm{R}$-operator on all  parameters.

Let us emphasize that the general $\mathrm{R}$-operator is the same in (\ref{RLL1}) and (\ref{RLL2}).
As we will see shortly it is invariant under $\omega \rightleftarrows \omega'$.

Further we propose two constructions of the $\mathrm{R}$-operator.

\subsection{The first construction}

The equation~(\ref{RLL1}) admits a natural interpretation: the
$\mathrm{R}$-operator interchanges the set of
parameters $(u_1,u_2)$ in the first $\mathrm{L}$-operator with
the set of parameters $(v_1,v_2)$ in the second
$\mathrm{L}$-operator. The operator $\mathrm{R}(\mathbf{u})$
corresponds to a particular permutation $s$ in the group of
permutations of four parameters $\mathfrak{S}_4$:
$$
s \rightarrow \mathrm{R}(\mathbf{u})\ ;\
s\,\mathbf{u}\equiv s\,(u_2,u_1,v_2,v_1)=(v_2,v_1,u_2,u_1).
$$
Any permutation from the group $\mathfrak{S}_4$ can be
composed from the elementary transpositions $s_{1}$,
$s_{2}$, and $s_{3}$:
$$
s_{1}\mathbf{u} = (u_1,u_2,v_2,v_1)\ , \
s_{2}\mathbf{u}
 = (u_2,v_2,u_1,v_1) \ , \
s_{3}\mathbf{u} = (u_2,u_1,v_1,v_2)\,,
$$
which interchange only two nearest neighboring elements
$$
(\overset{s_1}{\overbrace{u_2\,, u_1}}\,,
\overset{s_{3}}{\overbrace{v_2\,, v_1}})\ ;\ (u_2\
,\overset{s_2}{\overbrace{u_1\,, v_2}}\,, v_1)\,.
$$
It is natural to
search for the operators representing these elementary
transpositions in L-operators.
Namely, we demand that $\mathrm{S}_i(\mathbf{u})$ obey the following defining relations
\begin{gather}\label{RLL13} 
\mathrm{S}_1(\mathbf{u})\,\mathrm{L}_1(u_1,u_2) =
\mathrm{L}_1(u_2,u_1)\,\mathrm{S}_1(\mathbf{u})\ ;\
\mathrm{S}_3(\mathbf{u})\,\mathrm{L}_2(v_1,v_2) =
\mathrm{L}_2(v_2,v_1)\,\mathrm{S}_3(\mathbf{u}),
\\[0.2 cm] \label{RLL22}  
\mathrm{S}_2(\mathbf{u})\,\mathrm{L}_1(u_1,u_2)\,\mathrm{L}_2(v_1,v_2)=
\mathrm{L}_1(v_2,u_2)\,\mathrm{L}_2(v_1,u_1)\,\mathrm{S}_2(\mathbf{u})\,.
\end{gather}
In the previous Section we have already seen these relations. Indeed (\ref{WL2}) leads to
$$
D_{u_2-u_1}(p_1)\,\mathrm{L}_1(u_1,u_2) =
\mathrm{L}_1(u_2,u_1)\,D_{u_2-u_1}(p_1)\ ;\
D_{v_2-v_1}(p_2)\,\mathrm{L}_2(v_1,v_2) =
\mathrm{L}_2(v_2,v_1)\,D_{v_2-v_1}(p_2)\,
$$
and (\ref{SLL}) can be reformulated as
$$
D_{u_1-v_2}(x_{12})\,\mathrm{L}_1(u_1,u_2)\,\mathrm{L}_2(v_1,v_2) =
\mathrm{L}_1(v_2,u_2)\,\mathrm{L}_2(v_1,u_1)\,D_{u_1-v_2}(x_{12})\,,
$$
so that we have the following identification
$$
\mathrm{S}_1(\mathbf{u}) = D_{u_2-u_1}(p_1)\ ;\
\mathrm{S}_2(\mathbf{u}) = D_{u_1-v_2}(x_{12})\ ;\
\mathrm{S}_3(\mathbf{u}) = D_{v_2-v_1}(p_2)\,.
$$
The operator $\mathrm{R}(\mathbf{u})$
corresponds to a particular permutation $s = s_2 s_1s_3s_2$ in the group $\mathfrak{S}_4$. We have the following correspondence between permutations and operators
\begin{equation}
s_i \longrightarrow \mathrm{S}_i(\mathbf{u})\ ;\quad s_i s_j \longrightarrow
\mathrm{S}_i(s_j\mathbf{u})\,\mathrm{S}_j(\mathbf{u})\ ;\quad s_i s_j s_k \longrightarrow
\mathrm{S}_i(s_js_k\mathbf{u})\,\mathrm{S}_j(s_k\mathbf{u})
\,\mathrm{S}_k(\mathbf{u})\ \ldots \ \,,
\end{equation}
and it is easy to see that the composite operator
$$
\mathrm{R}(\mathbf{u}) =
\mathrm{S}_2(s_1s_3s_2\mathbf{u})\,
\mathrm{S}_1(s_3s_2\mathbf{u})\,\mathrm{S}_3(s_2\mathbf{u})\,
\mathrm{S}_2(\mathbf{u})
$$
satisfies equation~(\ref{RLL1}). In explicit form we have
\be \lb{R}
\mathrm{R}(\mathbf{u}) =
D_{u_2 - v_1}(x_{12}) \, D_{u_1 - v_1}(p_2)\,
D_{u_2 - v_2}(p_1)\, D_{u_1 - v_2}(x_{12})\,.
\ee
This operator is invariant under $\omega \rightleftarrows \omega'$ so that
it solves both $\mathrm{R}\mathrm{LL}$-relations (\ref{RLL1}), (\ref{RLL2}).

Let us rewrite this expression using spectral parameter and initial spin parameters.
The $\mathrm{R}$-operator depends only on the difference of spectral parameters $u-v$
so that taking into account (\ref{lambdaparam}) and substituting $u-v \to u$ one obtains
\be \lb{Rop}
\mathrm{R}_{12}(u) = D_{u-\nicefrac{(s_1+s_2)}{2}}(x_{12}) \, D_{u-\nicefrac{(s_1-s_2)}{2}}(p_2) \,
D_{u+\nicefrac{(s_1-s_2)}{2}}(p_1) \, D_{u+\nicefrac{(s_1+s_2)}{2}}(x_{12})\,.
\ee
In view of (\ref{Wint}) $\mathrm{R}$-operator can also be rewritten
as an integral operator
\be \lb{Rint}
\mathrm{R}_{12}(u) \, \Phi(x_1,x_2) =
\int \mathrm{d} x'_1 \mathrm{d} x'_2 \,
D_{u-\nicefrac{(s_1+s_2)}{2}}(x_1-x_2) \,D_{-\omega''-u+\nicefrac{(s_1-s_2)}{2}}(x_2-x'_2)\cdot
\ee
$$
\cdot D_{-\omega'' - u-\nicefrac{(s_1-s_2)}{2}}(x_1-x'_1) \,D_{u+\nicefrac{(s_1+s_2)}{2}}(x'_1-x'_2)\,
\Phi(x'_1,x'_2)\,.
$$
Note that we use a common notation $\mathrm{R}_{12}(u)$ for $\mathrm{R}$-operator.
Subscripts $1,2$ denote quantum spaces where the operator acts nontrivially,
and we indicate only the dependence on the spectral parameter $u$ skipping all other parameters for simplicity.
The $\mathrm{R}$-operator can be represented in two forms
which are complementary to each other: as a function of
operator argument~(\ref{Rop}) and as an integral operator~(\ref{Rint}).

We have constructed operator $\mathrm{R}_{12}(u)$ which solves
$\mathrm{R}\mathrm{LL}$-relations (\ref{RLL1}), (\ref{RLL2}) and now we have to check that corresponding operator with permutation
$\mathbb{R}_{12}(u)=\mathbb{P}_{12}\,\mathrm{R}_{12}(u)$ respects the Yang-Baxter equation
$$
\mathbb{R}_{12}(u - v) \, \mathbb{R}_{13}(u) \,  \mathbb{R}_{23}(v) = \mathbb{R}_{23}(v) \, \mathbb{R}_{13}(u)  \,  \mathbb{R}_{12}(u - v)
$$
in the tensor product $\pi_{s_1}\otimes\pi_{s_2}\otimes\pi_{s_3}$.
We state that if two following Coxeter relations take place
\begin{eqnarray}\label{def1}
s_1 s_2 s_1 = s_2 s_1 s_2 \longrightarrow
\mathrm{S}_1(s_2s_1\mathbf{u})\,\mathrm{S}_2(s_1\mathbf{u})\,
\mathrm{S}_1(\mathbf{u})=
\mathrm{S}_2(s_1s_2\mathbf{u})\,\mathrm{S}_1(s_2\mathbf{u})\,
\mathrm{S}_2(\mathbf{u})\,,
\\ [0.2 cm]\label{def3}
s_2 s_3 s_2 = s_3 s_2 s_3 \longrightarrow
\mathrm{S}_2(s_3s_2\mathbf{u})\,\mathrm{S}_3(s_2\mathbf{u})\,
\mathrm{S}_2(\mathbf{u})=
\mathrm{S}_3(s_2s_3\mathbf{u})\,\mathrm{S}_2(s_3\mathbf{u})\,
\mathrm{S}_3(\mathbf{u})\,,
\end{eqnarray}
then the Yang-Baxter equation is satisfied. For more details see~\cite{DKK,DM,DS}.

Relations~(\ref{def1}) and~(\ref{def3}) are equivalent to the relations
\be \lb{WSW1}
D_{u}(p_{1})\,D_{u+v}(x_{12})\,D_{v}(p_{1}) =
D_{v}(x_{12})\,D_{u+v}(p_{1})\,D_{u}(x_{12})\,,
\ee
\be \lb{WSW2}
D_{u}(p_{2})\,D_{u+v}(x_{12})\,D_{v}(p_{2}) =
D_{v}(x_{12})\,D_{u+v}(p_{2})\,D_{u}(x_{12})\,\,\,
\ee
which are reduced to the known star-triangle relation~\cite{VF,V}
\be \label{star-triang}
D_{u} (p) \; D_{u+v}(x) \; D_{v} (p) =
D_{v}(x) \; D_{u+v}(p) \; D_{u} (x)\,.
\ee
This relation can be rewritten in the integral form due to (\ref{Wint})
$$
\frac{A(u)A(v)}{A(u+v)} \int\limits^{+\infty}_{-\infty} \mathrm{d} x'' \;
D_{-\omega''-u}(x-x'') \, D_{u+v} (x'') \,
D_{-\omega''-v}(x''-x') = D_{v}(x) \, D_{-\omega''-u-v}(x-x') \, D_{u} (x')\,
$$
and it is a special example of the integral
relation~\cite{BT1,BMS} for the $D$-functions
$$
A(a) \,A(b)\, A(c) \int\limits^{+\infty}_{-\infty} D_a(z-z_1) D_b(z-z_2) D_c(z-z_3) =
$$
\be \lb{str-trg}
= D_{-\omega'' - a}(z_2-z_3) \, D_{-\omega'' - b}(z_3-z_1)\,D_{- \omega'' - c}(z_1-z_2) \;\;\; \mbox{at} \;\; a+b+c = -2\omega''\,.
\ee
The generalization of Faddeev-Volkov type solution of star-triangle relation is discussed in~\cite{Sp}.

\subsection{The second construction}
\label{sec-sec-con}

In this section we shall reproduce the solution of the
$\mathrm{RLL}$-relation using a different approach based on
the factorization of the $\mathrm{L}$-operator in terms of  $\mathrm{L}^{\pm}$-operators~(\ref{fL-L+}), (\ref{fL+L-}).

The initial $\mathrm{RLL}$-relation for the operator $\mathrm{R}(\mathbf{u})$
$$
\mathrm{R}(\mathbf{u}) \,\mathrm{L}_1(u_1,u_2)\,
\mathrm{L}_2(v_1,v_2) =
\mathrm{L}_1(v_1,v_2)\, \mathrm{L}_2(u_1,u_2)\, \mathrm{R}(\mathbf{u})
$$
is equivalent to the relation
\be \lb{L-L+L-L+'}
\mathrm{R}^{ab}(\mathbf{u})\,\mathrm{L}^{-}_a(u_2)
\,\mathrm{L}^{+}_1(u_1)\,\mathrm{L}^{-}_2(v_2)\,\mathrm{L}^{+}_b(v_1) =
\mathrm{L}^{-}_a(v_2)\,\mathrm{L}^{+}_1(v_1)\,\mathrm{L}^{-}_2(u_2)
\,\mathrm{L}^{+}_b(u_1)\,\mathrm{R}^{ab}(\mathbf{u})\,,
\ee
for the operator $\mathrm{R}^{ab}(\mathbf{u})$ which is obtained from
$\mathrm{R}(\mathbf{u})$ by similarity transformation
$$
\mathrm{R}^{ab}(\mathbf{u}) =
e^{i \pi (p_b - x_{2b})^2} e^{-i \pi (p_a - x_{a1})^2}\,
\mathrm{R}(\mathbf{u})\,
e^{i \pi (p_a - x_{a1})^2} e^{-i \pi (p_b - x_{2b})^2}\,.
$$
Here $x_{a1} = x_a-x_1$, $x_{2b} = x_2-x_b$ and the operators $x,p$ entering
$\mathrm{L}^{\pm}_a$ and $\mathrm{L}^{\pm}_b$ are replaced
by $x_a\,,p_a$ and $x_b\,,p_b$, respectively.

To transform the initial $\mathrm{RLL}$-relation to the
form~(\ref{L-L+L-L+'}) we multiply it by simple diagonal matrix
$
diag\left(e^{-\frac{i \pi}{2 \omega} p_a}\,,
e^{\frac{i \pi}{2 \omega} p_a}\right)$ from the left hand side and by
$
diag\left(
e^{-\frac{i \pi}{2 \omega} p_b}\,,
e^{\frac{i \pi}{2 \omega} p_b}\right)$ from the right hand side and take into account local factorization (\ref{fL-L+})-(\ref{fL+L-}).

Using the interchange relations (\ref{intwL-L+}) and (\ref{intwL+L-}) it is easy to check that operator
$$
\mathrm{R}^{ab}(\mathbf{u}) = D_{u_2-v_1}(x_{12})\,D_{u_2-v_2}(p_a+p_1-x_{a1})
\,D_{u_1-v_1}(p_2+p_b-x_{2b})\,D_{u_1-v_2}(x_{12})\,
$$
solves the relation (\ref{L-L+L-L+'}).
Then after needed similarity transformation we reproduce the expression for the operator $\mathrm{R}(\mathbf{u})$ from the previous Section
$$
\mathrm{R}(\mathbf{u}) =
e^{i \pi (p_a - x_{a1})^2}\,e^{-i \pi (p_b - x_{2b})^2}\,
\mathrm{R}^{ab}(\mathbf{u})
\,e^{i \pi (p_b - x_{2b})^2}\,e^{-i \pi (p_a - x_{a1})^2} =
$$
$$
= D_{u_2-v_1}(x_{12})\,D_{u_2-v_2}(p_1)\,D_{u_1-v_1}(p_2)\,D_{u_1-v_2}(x_{12})\,.
$$
This construction is based on two local factorization formulae (\ref{fL-L+}), (\ref{fL+L-}),
connection between two reduced $\mathrm{L}$-operators (\ref{L+toL-})
and the expression for intertwining operator (\ref{Wdiff}).

Note that the first construction of $\mathrm{R}$-operator
is the same as for elliptic modular double~\cite{DS}.
In principle all needed formulae can be obtained by some
reduction from the elliptic case but the direct
approach is much simpler.

We do not know the analog of the second construction
in elliptic situation. In Appendix we
show for completeness how it works for the group $\mathrm{SL}(2,\C)$.

\section{Reductions of the general $\mathrm{R}$-operator}

In this section we show that universal R-matrix $\mathcal{R}_{12}$ and its Yang-Baxterized version $\mathcal{R}_{12}(u)$ can be extracted from the constructed R-operator.
Let us perform certain reductions of the $\mathrm{R}$-operator with permutation $\mathbb{R}_{12}(u) = \mathbb{P}_{12}\,\mathrm{R}_{12}(u)$ where $\mathrm{R}_{12}(u)$ is given by~(\ref{Rop}).

Using asymptotic behaviour: $\gamma(z)\to 1$ for
$\Re(z)\to +\infty$ and reflection formula~(\ref{refl}) we obtain
\be \lb{red}
e^{4 \pi i (u + v)^2} \, e^{2 \pi i v p_1} \,
\mathbb{R}_{12}(u + v) \, e^{- 2 \pi i v p_1}
\to \mathcal{R}_{12}(u) \;\;\;\; \mbox{at} \;\; v \to +\infty
\ee
where
$$
\mathcal{R}_{12}(u) = e^{4 \pi i u^2}\,\mathbb{P}_{12}\,
\frac{e^{-2 \pi i (u+\frac{s_1+s_2}{2})x_{12}}}
                                     {\gamma(-x_{12}+\frac{s_1+s_2}{2}+u)}
\frac{e^{2 \pi i (u+\frac{s_2-s_1}{2})p_{2} }}
                                     {\gamma(p_{2}+\frac{s_2-s_1}{2}+u)}
\frac{e^{-2 \pi i (u+\frac{s_1-s_2}{2}) p_{1} }}
                                     {\gamma(-p_{1}+\frac{s_1-s_2}{2}+u)}
\frac{e^{2 \pi i (u -\frac{s_1+s_2}{2})x_{12}}}
                                     {\gamma(x_{12}-\frac{s_1+s_2}{2}+u)}\,.
$$
It is easy to check that dependence on the spectral parameter is defined by a simple similarity transformation
\be \lb{rBaxt}
\mathcal{R}_{12}(u) =
e^{- 2 \pi i u p_1} \, \mathcal{R}_{12} \,
e^{2 \pi i u p_1}\,,
\ee
where
\be \lb{r}
\mathcal{R}_{12} = \mathbb{P}_{12}\,
\frac{e^{-\pi i(s_1+s_2)x_{12}}}
{\gamma(-x_{12}+\frac{s_1+s_2}{2})}
\frac{e^{\pi i (s_2-s_1)p_{2} }}
{\gamma(p_{2}+\frac{s_2-s_1}{2})}
\frac{e^{- \pi i (s_1-s_2) p_{1} }}
{\gamma(-p_{1}+\frac{s_1-s_2}{2})}
\frac{e^{-\pi i (s_1+s_2)x_{12}}}
{\gamma(x_{12}-\frac{s_1+s_2}{2})}\,.
\ee
The operator $\mathcal{R}_{12}$ respects Yang-Baxter
equation without spectral parameters
\begin{equation}\label{YB0}
\mathcal{R}_{23}\,\mathcal{R}_{13}\,
\mathcal{R}_{12} =
\mathcal{R}_{12}\,
\mathcal{R}_{13}\,\mathcal{R}_{23}\,.
\end{equation}
As a direct consequence of this relation we immediately
obtain the Yang-Baxter equation
$$
\mathcal{R}_{23}(v)\,\mathcal{R}_{13}(u)\,
\mathcal{R}_{12}(u-v)=
\mathcal{R}_{12}(u-v)\,
\mathcal{R}_{13}(u)\,\mathcal{R}_{23}(v)\,
$$
for the operator $\mathcal{R}_{12}(u)$~(\ref{rBaxt}).

The operator (\ref{r}) plays the role of the universal R-matrix
for the modular double. In this form it
first appeared in the paper~\cite{BT} where it has been obtained
after some nontrivial calculations starting with
expression for universal $\mathrm{R}$-matrix in terms of quantum algebra generators~\cite{F}.

\medskip

The simplest way to prove~(\ref{YB0}) is to start from the
Yang-Baxter equation
\be \lb{YB1}
\mathbb{R}_{23}(v)\,\mathbb{R}_{13}(u)\,
\mathbb{R}_{12}(u-v)=
\mathbb{R}_{12}(u-v)\,
\mathbb{R}_{13}(u)\,\mathbb{R}_{23}(v)
\ee
and to perform the appropriate reductions. We have
$$
\mathbb{R}_{23}(v)\cdot
\underline{e^{2\pi i u p_1}\,\mathbb{R}_{13}(u) \, e^{-2\pi i u p_1}} \,
e^{2\pi i v p_1}
\,
\underline{e^{2\pi i (u-v) p_1}\, \mathbb{R}_{12}(u-v) e^{-2\pi i (u-v) p_1}}
\, e^{-2\pi i v p_1} =
$$
$$
= e^{2\pi i v p_1}\,
\underline{e^{2\pi i (u - v) p_1} \, \mathbb{R}_{12}(u-v)
\,e^{-2\pi i (u - v) p_1}} \, e^{-2\pi i v p_1} \,
\underline{e^{2\pi i u p_1} \,
\mathbb{R}_{13}(u)\, e^{-2\pi i u p_1}} \, \mathbb{R}_{23}(v)\,,
$$
so that it is possible to take the limit $u \to +\infty$ in
the underlined factors by means of (\ref{red}).
In view of translation invariance of the operator (\ref{r}):  $[\mathcal{R}_{12},p_1+p_2]=0$, one obtains
$$
\underline{e^{2\pi i v p_2}\, \mathbb{R}_{23}(v)\, e^{-2\pi i v p_2}}
\, \mathcal{R}_{13}\,
\mathcal{R}_{12} =
\mathcal{R}_{12}\,
\mathcal{R}_{13}\,
\underline{e^{2\pi i v p_2}\,\mathbb{R}_{23}(v)\,e^{-2\pi i v p_2}}\,,
$$
and taking the limit $v \to \infty$ in the underlined factors we obtain the Yang-Baxter relation~(\ref{YB0}).

\medskip

Note that there exists the additional three-term relation
\begin{equation}\label{Rrr}
\mathbb{R}_{23}(v)\,\mathcal{R}_{13}(u)\,
\mathcal{R}_{12}(u-v)=
\mathcal{R}_{12}(u-v)\,
\mathcal{R}_{13}(u)\,\mathbb{R}_{23}(v)\,,
\end{equation}
and corresponding analog of the $\mathrm{RLL}$-relation
\bea \lb{rlL}
\mathcal{R}_{12}(u-v) \, \ell_1(u) \, \mathrm{L}_2(v) =
 \mathrm{L}_2(v)\, \ell_1(u) \, \mathcal{R}_{12}(u-v)\,,
\\ [0.2 cm] \lb{rLl}
\mathcal{R}_{12}(u-v) \, \mathrm{L}_1(u)  \, \overline{\ell}_2(v)=
 \overline{\ell}_2(v) \, \mathrm{L}_1(u)\, \mathcal{R}_{12}(u-v)\,,
\eea
where operators $\ell,\,\overline{\ell}$ are obtained by some
reduction from the standard $\mathrm{L}$-operator
\be \lb{lbarl}
\begin{array}{c}
e^{-\frac{i \pi}{\omega}v }\, e^{2\pi i p v}\,
\mathrm{L}(u+v)\, e^{-2\pi i p v} \to
\ell(u) \equiv
\begin{pmatrix}
e^{\frac{i \pi}{\omega} u} \,\mathbf{K}_s & 0 \\
\mathbf{e}_s & e^{\frac{i \pi}{\omega} u} \,\mathbf{K}^{-1}_s
\end{pmatrix}\,, \;\;\;\; \text{for} \;\; v \to +\infty\,,
\\[0.2 cm]
- e^{\frac{i \pi}{\omega}v }
\, e^{2\pi i p v}
\, \mathrm{L}(u+v)\, e^{-2\pi i p v} \to
\overline{\ell}(u) \equiv
\begin{pmatrix}
e^{-\frac{i \pi}{\omega} u} \,\mathbf{K}^{-1}_s & -\mathbf{f}_s \\
0 & e^{-\frac{i \pi}{\omega} u} \,\mathbf{K}_s
\end{pmatrix}\,, \;\;\;\; \text{for} \;\; v \to -\infty\,.
\end{array}
\ee

To derive~(\ref{Rrr}) we again start with the Yang-Baxter
equation (\ref{YB1}) and transform it as follows
$$
\mathbb{R}_{23}(v)\,
\underline{e^{2 \pi i w p_1}\,\mathbb{R}_{13}(u+w)\, e^{-2 \pi i w p_1}} \,
\underline{e^{2 \pi i w p_1} \, \mathbb{R}_{12}(u-v+w) e^{-2 \pi i w p_1}} =
$$
$$
=\underline{e^{2 \pi i w p_1} \, \mathbb{R}_{12}(u-v+w)\, e^{-2 \pi i w p_1}} \,
\underline{e^{2 \pi i w p_1} \, \mathbb{R}_{13}(u+w)\,e^{-2 \pi i w p_1}} \,
\mathbb{R}_{23}(v)\,.
$$
Then we take the limit $w \to \infty$ in the underlined factors and obtain~(\ref{Rrr}).

The formulae~(\ref{rlL}) and~(\ref{rLl}) are consequence of the standard
$\mathrm{RLL}$-relation
$$
\mathbb{R}_{12}(u-v)\,\mathrm{L}_1(u)\,\mathrm{L}_2(v)=
\mathrm{L}_2(v)\,\mathrm{L}_1(u)\,\mathbb{R}_{12}(u-v)\,.
$$
For example, to derive~(\ref{rlL}) we transform it to the form
$$
\underline{e^{2 \pi i w p_1}\,\mathbb{R}_{12}(u-v+w)\,e^{-2 \pi i w p_1}} \,
\underline{ e^{2 \pi i w p_1}\,\mathrm{L}_1(u+w)\,e^{-2 \pi i w p_1} } \, \mathrm{L}_2(v) =
$$
$$
= \mathrm{L}_2(v) \cdot \underline{ e^{2 \pi i w p_1} \,\mathrm{L}_1(u+w)\, e^{-2 \pi i w p_1} } \cdot
\underline{e^{2 \pi i w p_1}\,\mathbb{R}_{12}(u-v+w)\,e^{-2 \pi i w p_1}}
$$
and take the limit $w \to +\infty$ using evident formulae for similarity transformation of the generators in representation (\ref{Gs})
$$
e^{2 \pi i u p} \, \mathbf{e}_s \, e^{-2 \pi i u p} = e^{ \frac{\pi i}{\omega} u } \, \mathbf{e}_s \;,\;\;\;
e^{2 \pi i u p} \, \mathbf{f}_s \, e^{-2 \pi i u p} = e^{- \frac{\pi i}{\omega} u } \, \mathbf{f}_s\,.
$$

\section*{Acknowledgement}

We thank A.Bytsko for discussions and critical remarks.
The work of D.~C. is supported by the Chebyshev Laboratory
(Department of Mathematics and Mechanics, St.-Petersburg State University)
under RF government grant 11.G34.31.0026,
and by Dmitry Zimin's "Dynasty" Foundation.
The work of S.~D. is supported by RFBR grants
11-01-00570,12-02-91052 and 13-01-12405.


\appendix
\renewcommand{\theequation}{\Alph{section}.\arabic{equation}}
\setcounter{table}{0}
\renewcommand{\thetable}{\Alph{table}}

\section*{Appendices}


\section{The $\mathbb{R}$-operator with $\mathrm{SL}(2,\mathbb{C})$-symmetry}
\setcounter{equation}{0}

In this Appendix we adopt the $\mathrm{R}$-operator construction of the Subsection~\ref{sec-sec-con}
for $\mathrm{SL}(2,\C)$ group.
We start with holomorphic
$\mathrm{L}$-operator
$$
\mathrm{L}(u_{1},u_{2}) = \left(%
\begin{array}{cc}
  1 & 0 \\
  z & 1 \\
\end{array}%
\right)\left(%
\begin{array}{cc}
  u_1 & -\partial \\
  0 & u_2 \\
\end{array}%
\right)\left(%
\begin{array}{cc}
  1 & 0 \\
  -z & 1 \\
\end{array}%
\right)\,, 
$$
and performing the limits $u_1 \to \infty$ or $u_2 \to \infty$ we
produce a pair of reduced $\mathrm{L}$-operators
$$
\mathrm{L}^{+}(u_{1}) = \left(%
\begin{array}{cc}
  u_1 & -\partial \\
  0 & 1 \\
\end{array}%
\right)\left(%
\begin{array}{cc}
  1 & 0 \\
  -z & 1 \\
\end{array}%
\right)\;,\;\;\;
\mathrm{L}^{-}(u_{2}) =
\left(%
\begin{array}{cc}
  1 & 0 \\
  z & 1 \\
\end{array}%
\right)
\left(%
\begin{array}{cc}
  1 & -\partial \\
  0 & u_2 \\
\end{array}%
\right)\,.
$$
In the following we omit the antiholomorphic sector (for more details see~\cite{DM}).
It can be checked that local factorizations take place (cf. (\ref{fL-L+})-(\ref{fL+L-}))
\be \lb{f1}
\mathrm{L}^{-}_1(v)\,\mathrm{L}^{+}_2(u) =
e^{-z_1 \partial_2}\, \mathrm{L}_1(u,v)
\left(
\begin{array}{cc}
  1 & 0 \\
  -z_2 & 1 \\
\end{array}
\right) \, e^{z_1 \partial_2}\,,
\ee
\be \lb{f2}
\mathrm{L}^{-}_1(v)\,\mathrm{L}^{+}_2(u) =
e^{-z_2 \partial_1}\,
\left(
\begin{array}{cc}
  1 & 0 \\
  z_1 & 1 \\
\end{array}
\right)
\mathrm{L}_2(u,v)
\, e^{z_2 \partial_1}\,.
\ee
Then we take into account the intertwining relation (cf. (\ref{WL2}))
$$
\partial^{v-u}_1\,\mathrm{L}_1(u,v) = \mathrm{L}_1(v,u)\,\partial^{v-u}_1\,,
$$
multiply it by the matrix
$\left(
\begin{array}{cc}
  1 & 0 \\
  -z_2 & 1 \\
\end{array}
\right)$ on the right, apply (\ref{f1}) and
$e^{-z_1\partial_2} \partial_1^{v-u} e^{z_1\partial_2} = (\partial_1+\partial_2)^{v-u}$
that gives the intertwining relation for $\mathrm{L}^{-}_1\,\mathrm{L}^{+}_2$ (cf. (\ref{intwL-L+})),
\be \lb{L-L+}
(\partial_1+\partial_2)^{v-u} \,
\mathrm{L}^{-}_1(v)\,\mathrm{L}^{+}_2(u)
=
\mathrm{L}^{-}_1(u)\,\mathrm{L}^{+}_2(v) \,
(\partial_1+\partial_2)^{v-u}\,.
\ee
To obtain a similar intertwining relation for $\mathrm{L}^{+}_1\,\mathrm{L}^{-}_2$
we note that the composition of the matrix similarity transformation and the canonical transformation
connects
$\mathrm{L}^{+}$ and $\mathrm{L}^{-}$ (cf. \ref{L+toL-}),
$$
\left.
\left(
\begin{array}{cc}
0 & 1 \\
1 & 0
\end{array}
\right)
\mathrm{L}^{-}(u)
\left(
\begin{array}{cc}
0 & 1 \\
1 & 0
\end{array}
\right) \right|_{z \to -\partial,\, \partial \to z} =
\mathrm{L}^{+}(u)
\;,\;\;\;
\left.
\left(
\begin{array}{cc}
0 & 1 \\
1 & 0
\end{array}
\right)
\mathrm{L}^{+}(u)
\left(
\begin{array}{cc}
0 & 1 \\
1 & 0
\end{array}
\right) \right|_{z \to \partial,\, \partial \to -z} =
\mathrm{L}^{-}(u)\,,
$$
that leads to (cf. (\ref{intwL+L-}))
\be \lb{L+L-}
z_{12}^{v-u} \,
\mathrm{L}^{+}_1(v)\,\mathrm{L}^{-}_2(u)
=
\mathrm{L}^{+}_1(u)\,\mathrm{L}^{-}_2(v) \,
z_{12}^{v-u}\,.
\ee
Our aim is to construct the operator $\mathrm{R}(\mathbf{u})$
which solves $\mathrm{RLL}$-relation
$$
\mathrm{R}(\mathbf{u}) \,\mathrm{L}_1(u_1,u_2)\,
\mathrm{L}_2(v_1,v_2) =
\mathrm{L}_1(v_1,v_2)\, \mathrm{L}_2(u_1,u_2)\, \mathrm{R}(\mathbf{u})\,.
$$
We multiply it by
$\left(
\begin{array}{cc}
  1 & 0 \\
  z_a & 1 \\
\end{array}
\right)$ on the left and by
$\left(
\begin{array}{cc}
  1 & 0 \\
  -z_b & 1 \\
\end{array}
\right)$ on the right, substitute (\ref{f1}) and (\ref{f2})
that leads to
\be \lb{L-L+L-L+}
\mathrm{R}^{ab}(\mathbf{u})\,\mathrm{L}^{-}_a(u_2)\,\mathrm{L}^{+}_1(u_1)\,\mathrm{L}^{-}_2(v_2)\,\mathrm{L}^{+}_b(v_1) =
\mathrm{L}^{-}_a(v_2)\,\mathrm{L}^{+}_1(v_1)\,\mathrm{L}^{-}_2(u_2)\,\mathrm{L}^{+}_b(u_1)\,\mathrm{R}^{ab}(\mathbf{u})\,,
\ee
$$
\mathrm{R}^{ab}(\mathbf{u}) =
e^{-z_1 \partial_a} e^{ -z_2 \partial_b} \,\mathrm{R}(\mathbf{u})\, e^{z_1 \partial_a} e^{z_2 \partial_b}\,.
$$
We can easily solve (\ref{L-L+L-L+}) and find $\mathrm{R}^{ab}$
due to (\ref{L-L+}) and (\ref{L+L-}),
$$
\mathrm{R}^{ab}(\mathbf{u}) = z_{12}^{u_2-v_1} (\partial_a+\partial_1)^{u_2-v_2} (\partial_2+\partial_b)^{u_1-v_1} z_{12}^{u_1-v_2}\,.
$$
Applying the relation between $\mathrm{R}^{ab}(\mathbf{u})$ and $\mathrm{R}(\mathbf{u})$ we obtain (cf. (\ref{R}))
$$
\mathrm{R}_{12}(\mathbf{u}) =
z_{12}^{u_2-v_1} \partial_1^{u_2-v_2} \partial_2^{u_1-v_1} z_{12}^{u_1-v_2}\,.
$$
Taking into account the antiholomorohic sector the modification of the latter
$\mathrm{R}$-operator arises than can be rewritten in the integral form
which is analogous to (\ref{Rint})~\cite{DM}.


\end{document}